\documentclass[format=acmsmall, review=false, screen=true]{acmart}

\usepackage{booktabs} 

\usepackage[ruled]{algorithm2e} 
\usepackage{listings}
\usepackage{graphicx}
\usepackage{subcaption}
\definecolor{mygreen}{rgb}{0,0.6,0}
\definecolor{mygray}{rgb}{0.5,0.5,0.5}
\definecolor{mymauve}{rgb}{0.58,0,0.82}

\newcommand{\pysdc}{\texttt{pySDC}}
\newcommand{\matr}[1]{\mathbf{#1}}
\newcommand{\Qmat}{\matr{Q}}
\newcommand{\QDmat}{\matr{Q}_\Delta}
\newcommand{\vect}[1]{\boldsymbol{#1}}
\newcommand{\tvect}[1]{\vec{\vect{#1}}}
\newcommand{\dt}{\Delta t}

\acmJournal{TOMS}

\begin{document}
\title{pySDC -- Prototyping spectral deferred corrections} 

\author{Robert Speck}
\orcid{0000-0002-3879-1210}
\affiliation{%
  \institution{J\"ulich Supercomputing Centre, Forschungszentrum J\"ulich GmbH}
  \country{Germany}}
\email{r.speck@fz-juelich.de}
\renewcommand\shortauthors{Speck, R.}

\begin{abstract}
In this paper we present the Python framework \pysdc{} for solving collocation problems with spectral deferred correction methods (SDC) and their time-parallel variant PFASST, the parallel full approximation scheme in space and time.
\pysdc{} features many implementations of SDC and PFASST, from simple implicit time-stepping to high-order implicit-explicit or multi-implicit splitting and multi-level spectral deferred corrections.
It comes with many different, pre-implemented examples and has seven tutorials to help new users with their first steps.
Time-parallelism is implemented either in an emulated way for debugging and prototyping as well as using MPI for benchmarking.
The code is fully documented and tested using continuous integration, including most results of previous publications.
Here, we describe the structure of the code by taking two different perspectives: the user's and the developer's perspective.
While the first sheds light on the front-end, the examples and the tutorials, the second is used to describe the underlying implementation and the data structures.
We show three different examples to highlight various aspects of the implementation, the capabilities  and the usage of \pysdc{}.
Also, couplings to the FEniCS framework and PETSc, the latter including spatial parallelism with MPI, are described.
\end{abstract}

%
%


%
%


\maketitle

\section{Introduction}

For solving initial value problems, implicit integration methods based on collocation are an attractive approach, since they feature near-ideal or, depending on the choice of quadrature nodes, even ideal convergence orders and typically have very advantageous stability properties. 
However, solving the dense and fully coupled collocation problem directly is often prohibitively expensive, in particular for discretized partial differential equations: For $M$ collocation nodes and an $N$-dimensional system of ordinary differential equations (ODEs), a system of size $MN\times MN$ has to be solved. 
Here, an iterative strategy is favorable, where instead of the full system only $M$ smaller systems of size $N \times N$ need to be solved for each iteration.

Such an approach is given by the so-called ``spectral deferred correction method'' (SDC), introduced in~\cite{DuttEtAl2000}.
After rewriting the initial value problem into its Picard form, a provisional solution to the integral problem is computed using a standard time-stepping method, typically the explicit or the implicit Euler scheme. 
Then, this provisional solution is corrected using a sequence of error integral equations, which are also solved using one of the standard methods. 
This way, a higher-order time-stepping method can be obtained simply by using low-order methods repeatedly.
Xia et al.\ showed in~\cite{ShuEtAl2007} that each iteration or ``sweep'' of SDC can raise the order by one up to the order of the underlying collocation formula.
This has been further analyzed e.g. in~\cite{TangEtAl2013}, besides others.
In the last decade, SDC has been applied e.g.\ to gas dynamics and incompressible or reactive flows~\cite{BouzarthMinion2010,LaytonMinion2004,Minion2004} as well as to fast-wave slow-wave problems~\cite{RuprechtSpeck2016}, atmospheric modeling~\cite{JiaEtAl2013}, phase-field problems~\cite{FengEtAl2015_sdc} or particle motions in magnetic fields~\cite{WinkelEtAl2015}.
In addition to its flexibility, SDC has been proven to provide many opportunities for algorithmic and mathematical improvements, e.g.: 
\begin{itemize}
	\item convergence can be accelerated by GMRES~\cite{HuangEtAl2006} or algebraic preconditioners~\cite{Weiser2014},
    \item high-order implicit-explicit or even multi-implicit splitting is straightforward~\cite{Minion2003,RuprechtSpeck2016,BOURLIOUX2003651},
    \item multirate integration allows effective treatment of different time-scales~\cite{BouzarthMinion2010, NaumannEtAl2018}
    \item inexact spatial solvers enhance time-to-solution~\cite{10.1007/978-3-319-18827-0_39,WeiserGhosh2018}.
\end{itemize}

One of the key features of such an iterative approach for time-stepping, though, is that these approaches can be used to enable efficient parallel-in-time integration.
Using SDC, the ``parallel full approximations scheme in space and time'' (PFASST) by Emmett and Minion~\cite{EmmettMinion2012} allows to integrate multiple time-steps simultaneously by using SDC sweeps on a space-time hierarchy.
To this end, SDC makes use of multiple levels in space and time by using multigrid ideas~\cite{Speck2015,BoltenEtAl2017}.
This "multi-level spectral deferred corrections" is then extended to multiple concurrent time-steps, providing a multigrid-like algorithm for the so called composite collocation problem. 
PFASST has been applied to a variety of problems and coupled to different space-parallel solvers, ranging from a Barnes-Hut tree code to geometric multigrid, see e.g.~\cite{SpeckEtAl2012, MinionEtAl2015}. Together with spatial parallelization, PFASST was demonstrated to run and scale on up to 458,752 cores of an IBM Blue Gene/Q installation.
This ``parallelization across the steps'' approach~\cite{Burrage1997} targets large-scale parallelization on top of saturated spatial parallelization of partial differential equations (PDEs), where parallelization in the temporal domain acts as a multiplier for standard parallelization techniques in space.
The most prominent example of this variant of parallel-in-time integration methods is the Parareal algorithm by Lion, Maday and Turinici~\cite{LionsEtAl2001}.
Its success is accounted to its simplicity and applicability: Only a fine but expensive and a coarse but cheap propagator in time have to be provided by the user.
Then, parallelization across the temporal dimension can be achieved in an iterative prediction-correction manner. 
Parareal has sparked a lot of research and new results in the field of parallel-in-time integration and PFASST can be seen as one of its direct successors.

Besides parallelization of time-integration on multiple steps, ``parallelization across the method'' approaches~\cite{Burrage1997} aim at parallelizing the integration of each time-step individually.
While this typically results in small-scale parallelization in the time-domain, parallel efficiency and applicability of these methods are often more favorable.
Most notably, the ``revisionist integral deferred correction method'' (RIDC) by Christlieb et al.~\cite{ChristliebEtAl2010,Ong:2016:A9R:2987591.2964377} makes use of integral deferred corrections in order to compute multiple iterations in a pipelined way.
Also within the SDC context, parallelization across the method is possible: In~\cite{Speck2018}, two strategies for computing updates on multiple collocation nodes simultaneously is presented. 
This approach is related to parallel Runge-Kutta methods as presented e.g.\ in~\cite{VanderHouwen1990}. 
In~\cite{Gander2015_Review}, a much more structured and complete overview of parallel-in-time integration approaches can be found.
Also, the website \texttt{www.parallel-in-time.org} offers a comprehensive list of references and we refer to both resources for more details.

While many ideas, algorithms and proofs of concept exist in this domain, the number of accessible time-parallel application codes is actually quite small. 
Interestingly, the same is also true for stand-alone, more generic parallel-in-time libraries, which are mainly used for showcasing performance or for testing new ideas. 
In particular, codes which can deal with parallelization in time as well as in space are rare.
At the time of this writing, probably the most notable implementations are the following three:
\begin{description}
	\item[\texttt{XBraid}] - a C/C++ time-parallel multigrid solver (or, actually, more like a framework) generalizing the Parareal idea to multiple levels~\cite{xbraid},
    \item[\texttt{RIDC}] - a C++-implementation of the revisionist integral deferred correction method~\cite{Ong:2016:A9R:2987591.2964377}, 
    \item[PFASST] - different implementations of the parallel full approximation scheme in space and time~\cite{pfasst_code}.
\end{description}
Besides those libraries, there are of course many further implementations of parallel-in-time ideas, but which are either more focused, less maintained or not easily accessible/available.
In particular, we mention in this context the \texttt{SWEET} code~\cite{sweet}, various implementations of the PARAEXP method~\cite{doi:10.1137/110856137} and dependency-driven Parareal~\cite{Elwasif:2011:DFP:2132876.2132883}.

Note that for PFASST we did not cite a particular code but rather a website listing three different ones: the original Fortran library \texttt{libpfasst}, a specialized DUNE module \texttt{dune-PFASST} written in C++ and the Python implementation \pysdc{}. 
This paper is concerned with the latter of those implementations.
One may ask why there is need for yet another PFASST code, in particular with the rather generic \texttt{libpfasst} library already available.
As a user, starting with SDC, PFASST or parallel-in-time integration itself often corresponds to a quite severe investment of time, effort and endurance without any guarantee that the results will be worth it.
With \pysdc{} we provide a code which helps users to set up a prototype as fast and as easy as possible to see whether SDC, PFASST or even the collocation problem itself are the right way to go for her or his problem at hand.
The same is true for training students: a code which is easy to install and to use and which is close to the formulas in the relevant papers reduces the initial time it takes to start with the actual tasks.
Furthermore, for investigating new methods like variants of SDC or new coarse levels for PFASST, it is a relief to be able to ignore tedious implementation details, communication structures or lower-level language peculiarities.
With \pysdc{}, users as well as developers shall be enabled to focus on their own ideas and challenges.
It can be used to simply set up an ODE system and run standard versions of SDC or PFASST spending close to no thoughts on the internal structure.
It can also be used, however, to implement new iterative methods or to improve existing methods by overriding any component of \pysdc{}, from the main controller and the SDC sweeps to the transfer routines or the way the hierarchy is created.

Before we describe \pysdc{} and some of its components we start by introducing SDC and PFASST more thoroughly in the next section. 
Then, \pysdc{} is examined from a user's and a developer's perspective, highlighting different aspects of the implementation.
Finally, three numerical examples are given which demonstrate \pysdc{}'s capabilities.

\section{SDC and PFASST}

In this section we briefly review the collocation problem, being the basis for all problems the algorithm presented here tries to solve in one way or another. 
Then, spectral deferred corrections (SDC) are introduced, which then lead to the time-parallel variant PFASST, the "parallel full approximation scheme in space and time".
This section is largely based on~\cite{Speck2017,NLA:NLA2110}.

\subsection{Spectral deferred corrections}

For ease of notation we consider a scalar initial value problem
\begin{align*}
  u_t = f(u),\quad u(0) = u_0,
\end{align*}
with $u(t), u_0, f(u) \in\mathbb{R}$.
For an interval $[t_0,t_1]$, we rewrite this in Picard formulation as
\begin{align*}
  u(t) = u_0 + \int_{t_0}^t f(u(s))ds,\quad t\in[t_0,t_1].
\end{align*}
Introducing $M$ quadrature nodes $\tau_1,...,\tau_M$ with $t_l \le \tau_1 < ... < \tau_M = t_{l+1}$, we can approximate the integrals from $t_l$ to these nodes $\tau_m$ using spectral quadrature like Gauss-Radau or Gauss-Lobatto quadrature, such that
\begin{align*}
  u_m = u_0 + \sum_{j=1}^Mq_{m,j}f(u_j),\quad m = 1, ..., M,
\end{align*} 
where $u_m \approx u(\tau_m)$, $\dt = t_{1}-t_0$ and $q_{m,j}$ represent the quadrature weights for the interval $[t_0,\tau_m]$ with
\begin{align*}
  \sum_{j=1}^Mq_{m,j}f(u_j)\approx\int_{t_0}^{\tau_m}f(u(s))ds.
\end{align*}
We can now combine these $M$ equations into one system of linear or non-linear equations with
\begin{align}\label{eq:coll_prob}
  \left(\matr{I}_M - \dt\Qmat\vect{F}\right)(\vect{u}) = \vect{u}_0
\end{align}
where $\vect{u} = (u_1, ..., u_M)^T \approx (u(\tau_1), ..., u(\tau_M))^T\in\mathbb{R}^M$, $\vect{u}_0 = (u_0, ..., u_0)^T\in\mathbb{R}^M$, $\Qmat = (q_{i,j})_{i,j}\in\mathbb{R}^{M\times M}$ is the matrix gathering the quadrature weights, $\matr{I}_M$ is the identity matrix of dimension $M$ and the vector function $\vect{F}$ is given by $\vect{F}(\vect{u}) = (f(u_1), ..., f(u_M))^T\in\mathbb{R}^M$.
This system of equations is called the ``collocation problem'' and it is equivalent to a fully implicit Runge-Kutta method, where the matrix $\matr{Q}$ contains the entries of the corresponding Butcher tableau.
We note that for $f(u) \in\mathbb{R}^N$, we need to replace $\Qmat$ by $\Qmat\otimes\matr{I}_N$.

This system of equations is dense and a direct solution is not advisable, in particular if the right-hand side of the ODE is non-linear.
Using SDC, this problem can be solved iteratively and we follow~\cite{HuangEtAl2006,Weiser2014,RuprechtSpeck2016} to present SDC as preconditioned Picard iteration for the collocation problem~\eqref{eq:coll_prob}.
Standard discretized Picard iteration is given by
\begin{align*}
  \vect{u}^{k+1} = \vect{u}^{k} + \left(\vect{u}_0 - \left(\matr{I}_M - \dt\matr{Q}\vect{F}\right)(\vect{u}^k)\right)
\end{align*} 
for $k = 0, ... K$. 
This is simply an unmodified, non-linear Richardson iteration for \eqref{eq:coll_prob} and for very small $\dt$, this indeed converges to the solution of~\eqref{eq:coll_prob}.
In order to increase range and speed of convergence, we now precondition this iteration.
The standard approach to preconditioning is to define an operator which is easy to invert but also close to the operator of the system.
For SDC, we now choose a simpler quadrature rule for the preconditioner. 
In particular, the resulting matrix $\QDmat$ gathering the weights of this rule is a lower triangular matrix, such that solving the system can be easily done by forward substitution. 
We write
\begin{align}\label{eq:sdc_iteration}
  \left(\matr{I}_M - \dt\QDmat\vect{F}\right)(\vect{u}^{k+1}) = \vect{u}_0 + \dt(\Qmat-\QDmat)\vect{F}(\vect{u}^{k})
\end{align}
and the operator $\matr{I} - \dt\QDmat\vect{F}$ is then called the SDC preconditioner.
The matrix $\QDmat$ is typically given by the implicit Euler method which corresponds to the right-hand side rule in terms of integration 
or, using the LU decomposition of $\matr{Q}^T$~\cite{Weiser2014}, by
\begin{align*}
  \QDmat^{\mathrm{LU}} = \matr{U^T}\quad \text{for}\quad \Qmat^T = \matr{L}\matr{U}.
\end{align*}
This choice, which is also called the "LU trick", is very well suited for stiff problems and has become the de-facto standard choice for $\QDmat$.

In order to solve~\eqref{eq:sdc_iteration}, an SDC iteration proceeds from one collocation node to the next by solving
\begin{align}\label{eq:sdc_iteration_node}
	u^{k+1}_{m+1}-\dt \tilde{q}_{m+1,m+1}f(u^{k+1}_{m+1}) = u_0 + \dt\sum_{j=1}^m\tilde{q}_{m+1,j}f(u_j^{k+1}) + \dt\sum_{j=1}^M(q_{m+1,j} - \tilde{q}_{m+1,j})f(u_j^k),
\end{align}
for the $m+1$-th collocation node with $\QDmat = (\tilde{q}_{i,j})_{i,j}$, which simply corresponds to the $m+1$-th line of~\eqref{eq:sdc_iteration}. 
Due to the lower-triangular structure of $\QDmat$, $u_{m+1}^{k+1}$ only depends on the function values $f(u_0^{k+1}), ..., f(u_{m+1}^{k+1})$ as well as those at the previous iteration.
The matrix $\QDmat$ can also be chosen as strictly lower triangular matrix, so that the dependency of $f(u_{m+1}^{k+1})$ is removed and the SDC iteration becomes explicit.
A popular choice for such a $\QDmat$ is the left-hand side rule, which corresponds to the explicit Euler method.
However, if an implicit preconditioner is chosen, a "spatial solver" is required at each node to solve~\eqref{eq:sdc_iteration_node}.
This solver can be used as a blackbox, treating~\eqref{eq:sdc_iteration_node} as implicit Euler formula with modified right-hand side.
However, recent results indicate that the computational efficiency of SDC can be enhanced significantly, if the spatial solver is taken into account more carefully, see~\cite{10.1007/978-3-319-18827-0_39,WeiserGhosh2018}.

\subsection{Parallel full approximation scheme in space and time}

We can assemble the collocation problem~\eqref{eq:coll_prob} for multiple time-steps, too.
Let $\vect{u}_1, ..., \vect{u}_L$ be the solution vectors at time-steps $1,..., L$ and $\tvect{u} = \left(\vect{u}_1, ...,\vect{u}_L\right)^T$ the full solution vector. 
We define a matrix $\matr{H}\in\mathbb{R}^{M\times M}$ such that $\matr{H}\vect{u}_\ell$ provides the initial value for the $\ell+1$-th time-step.
Note that this initial value has to be used at all nodes, see the definition of $\vect{u}_0$ above.
The matrix depends on the collocation nodes and if the last node is the right interval boundary, i.e. $\tau_M=t_{\ell+1}$ as it is the case for Gauss-Radau or Gauss-Lobatto nodes, then it is simply given by 
\begin{align*}
	\matr{H} &=  
    \begin{pmatrix}
      0 & 0 & \cdots & 1 \\
      0 & 0 & \cdots & 1 \\
       \vdots & \vdots & & \vdots \\
      0 & 0 & \cdots & 1 \\
    \end{pmatrix}.
\end{align*}
Otherwise, $\matr{H}$ would contain weights for extrapolation or the collocation formula for the full interval.
Note that for $f(u) \in\mathbb{R}^N$, we again need to replace $\matr{H}$ by $\matr{H}\otimes\matr{I}_N$.
With this definition, we can assemble the so-called "composite collocation problem" for $L$ time-steps as
\begin{align}\label{eq:comp_coll_prob}
    \matr{C}(\tvect{u}) &:= 
    \begin{pmatrix}
      \matr{I}_M - \dt\Qmat\vect{F} &  & & \\
      -\matr{H}		 & \matr{I}_M - \dt\Qmat\vect{F} & & \\
                   & \ddots           & \ddots & \\
               &  &  -\matr{H} & \matr{I}_M - \dt\Qmat\vect{F}
    \end{pmatrix}
    \begin{pmatrix}
    	\vect{u}_1\\
        \vect{u}_2\\
        \vdots\\
        \vect{u}_L
    \end{pmatrix} = 
    \begin{pmatrix}
    	\vect{u}_0\\
        \vect{0}\\
        \vdots\\
        \vect{0}
    \end{pmatrix} =:
    \tvect{u}_0.
\end{align}
with $\tvect{u}_0 = \left(\vect{u}_0, \vect{0}, ..., \vect{0}\right)^T\in\mathbb{R}^{LM}$.
The operator $\matr{C}$ should be seen as non-linear operator acting on the vector $\tvect{u}$, if $f$ is non-linear.
If $f$ is linear, $\matr{C}\in\mathbb{R}^{LM\times LM}$ is simply a matrix.
More compactly, \eqref{eq:comp_coll_prob} can be written as
\begin{align}
	\left(\matr{I}_{LM} - \matr{I}_L\otimes\dt\Qmat\vect{F} - \matr{E}\otimes\matr{H}\right)(\tvect{u}) = \tvect{u}_0,
\end{align}
where the matrix $\matr{E}\in\mathbb{R}^{M\times M}$ has ones on the lower sub-diagonal and zeros elsewhere.

Using SDC for the collocation problem~\eqref{eq:coll_prob} gives us some options to solve the composite collocation problem~\eqref{eq:comp_coll_prob}.
In all of those options, SDC plays the role of a local inner iteration.
Each local iteration of SDC is then called a "sweep" and the actual SDC algorithm is the "sweeper".
The most obvious option for using SDC for the composite collocation problem is to solve each step after another, i.e. perform SDC sweeps until convergence on $\vect{u}_1$, move to step $2$ via $\matr{H}$, do SDC there and so on.
This corresponds to standard serial time-stepping.
Another serial, but more global approach is "serial multi-step SDC": here, we do one or more sweeps on a time-step and then send the results forward via $\matr{H}$ to the next one. 
Sweep there and move forward and so on until the last time-step is reached. Then repeat the process until all time-steps are converged.
This algorithm can be expressed in terms of a preconditioned iteration by choosing the preconditioner $\matr{P}_{\mathrm{ser}}$ via
\begin{align*}
	\matr{P}_{\mathrm{ser}}(\tvect{u}) &:= \left(\matr{I}_{LM} - \matr{I}_L\otimes\dt\QDmat\vect{F} - \matr{E}\otimes\matr{H}\right)(\tvect{u})
\end{align*}
and then solve the system
\begin{align*}
  \matr{P}_{\mathrm{ser}}(\tvect{u}^{k+1}) = \tvect{u}_0 + \left(\matr{P}_{\mathrm{ser}}-\matr{C}\right)(\tvect{u}^{k}).
\end{align*}
The operator $\matr{P}_{\mathrm{ser}}$ can be seen as approximative block Gauss-Seidel preconditioner, with $\QDmat$ replacing the original $\Qmat$ on the block diagonal.
Thus, although each time-step can be treated by a separate process, this is obviously a serial process.
In order to introduce parallelism over the time-steps, we introduce the approximative block Jacobi preconditioner with 
\begin{align*}
    \matr{P}_{\mathrm{par}}(\tvect{u}) := \left(\matr{I}_{LM} - \matr{I}_L\otimes\dt\QDmat\vect{F}\right)(\tvect{u}),
\end{align*}
where the lower block-diagonal and therefore the coupling between time-steps has been removed.
As before, we then solve the system
\begin{align*}
  \matr{P}_{\mathrm{par}}(\tvect{u}^{k+1}) = \tvect{u}_0 + \left(\matr{P}_{\mathrm{par}}-\matr{C}\right)(\tvect{u}^{k}),
\end{align*}
but now each time-step can be computed in parallel, no exchange of information happens within an iteration. 
Only when computing the term $\left(\matr{P}_{\mathrm{par}}-\matr{C}\right)(\tvect{u}^{k})$, information from previous time-steps is moved forward.
This "parallel multi-step SDC" of course leads to a severe degradation in convergence speed, requiring formally at least $L$ iterations to propagate information through the whole system.
We therefore have a fast converging, but serial method and a parallel, but slow converging one for solving the composite collocation problem.

The idea of PFASST now is to use these two approaches within a multigrid method.
More precisely, PFASST uses $\matr{P}_{\mathrm{par}}$ as smoother on fine levels and $\matr{P}_{\mathrm{ser}}$ as approximative solver on the coarsest level.
Restriction and prolongation operators $\matr{I}_h^H$ and $\matr{I}_H^h$ allow to transfer information between a fine level (indicated with $h$) and a coarse level (indicated with $H$).
The approximate solution is then used to correct the solution of the smoother on the finer level.
Typically, only two levels are used, although the method is not restricted to this choice.
As coarsening strategies, PFASST in its standard implementation allows to employ coarsening in the degrees-of-freedom in space (i.e. use $N/2$ instead of $N$ unknowns), a reduced collocation rule (i.e. use a different $\Qmat$ on the coarse level), a less accurate solver in space (for solving~\eqref{eq:sdc_iteration} on each time-step) or even a reduced representation of the problem.
The first two strategies directly influence the definition of the restriction and prolongation operators.

Since the right-hand side of the ODE can be a non-linear function, a "full approximation scheme" (FAS) is used, where a $\tau$-correction is added to the coarse problem. 
One PFASST iteration then works like this:
\begin{enumerate}
	\item compute $\tvect{u}_h^{k+1/2}$ from  
    \begin{align*}
    	\matr{P}_{\mathrm{par}}(\tvect{u}_h^{k+1/2}) = \tvect{u}_{0,h} + \left(\matr{P}_{\mathrm{par}}-\matr{C}_h\right)(\tvect{u}_h^{k})
    \end{align*}
    \item compute $\tau$-correction as 
    \begin{align*}
		\tvect{\tau} = \matr{C}_H\left(\matr{I}_h^H\tvect{u}^{k+1/2}_h\right) - \matr{I}_h^H\matr{C}_h\left(\tvect{u}^{k+1/2}_h\right)
	\end{align*}
    \item compute $\tvect{u}_H^{k+1}$ from 
    \begin{align*}
    	\matr{P}_{\mathrm{ser}}(\tvect{u}_H^{k+1}) = \tvect{u}_{0,H} + \tvect{\tau} + \left(\matr{P}_{\mathrm{ser}}-\matr{C}_H\right)(\matr{I}_h^H\tvect{u}_h^{k+1/2})
    \end{align*}
    \item compute $\tvect{u}_h^{k+1}$ from  
    \begin{align*}
    	\tvect{u}_h^{k+1} = \tvect{u}_h^{k+1/2} + \matr{I}_H^h\left(\tvect{u}_H^{k+1} - \matr{I}_h^H\tvect{u}_h^{k+1/2}\right)
    \end{align*}
\end{enumerate}
For more details on non-linear multigrid we refer to~\cite{trottenberg2000multigrid}.
We note that this "multigrid perspective" on PFASST~\cite{NLA:NLA2110} is rather new and does not represent the original idea of PFASST as described in~\cite{Minion2010,EmmettMinion2012}.
There, PFASST is presented as a coupling of SDC with the time-parallel method Parareal, augmented by the $\tau$-correction which allows to represent find-level information on the coarse level.
In~\cite{Speck2015}, then, the time-serial variant "multi-level spectral deferred corrections" has been analyzed.
By choosing $L=1$ in the derivations above, we obtain a multigrid method for the single-step collocation problem~\eqref{eq:coll_prob}, so that PFASST can also be seen as parallel version of MLSDC on multiple time-steps.
This understanding is indeed different from the multigrid perspective presented here: in the classical "parallel MLSDC" view, MLSDC is run on each time-step in parallel and quite independently, with data being sent forward on each level as needed.
In the novel multigrid view, all processes participate first in step (1), then (2) and so on. 

\begin{figure}[t]
  \centering
  \begin{subfigure}[b]{0.49\textwidth}
    \centering
    \includegraphics[width=0.95\columnwidth]{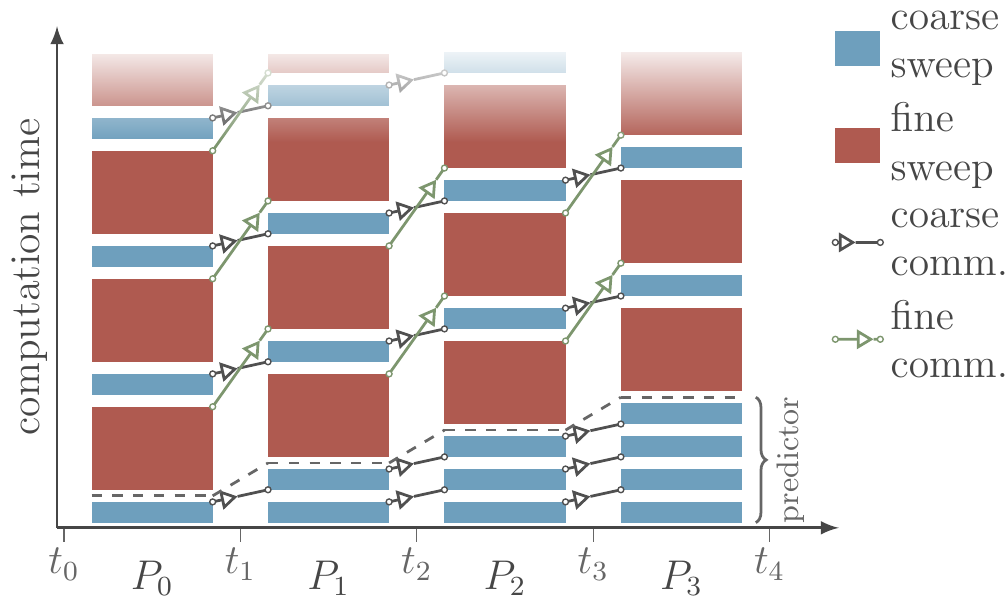}
    \caption{Classical implementation}
    \label{fig:pfasst_classic}
  \end{subfigure}
  \hfill
  \begin{subfigure}[b]{0.49\textwidth}
    \centering
    \includegraphics[width=0.95\columnwidth]{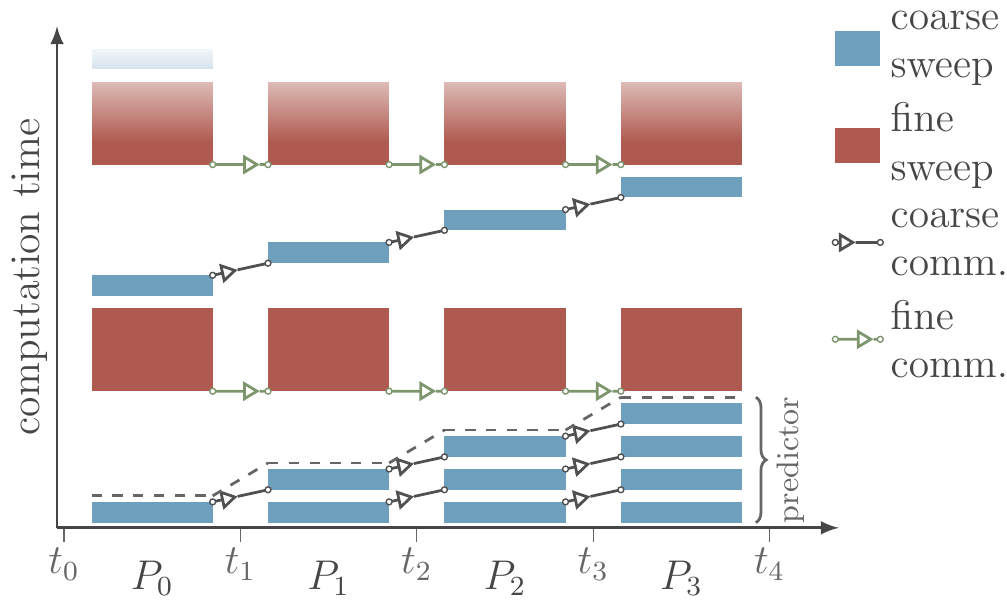}
    \caption{Multigrid implementation}
    \label{fig:pfasst_multigrid}
  \end{subfigure}
  \caption{Different implementation approaches for PFASST: \ref{fig:pfasst_classic} shows the classical, parallel MLSDC view, where on each process/time-step MLSDC is run, with fine communication being overlapped with coarse sweeps. \ref{fig:pfasst_multigrid}, in contrast, shows the multigrid view, with fine sweep ("smoothing step") clearly separated from the coarse sweep ("coarse-grid correction") and no overlapping communication.}
  \label{fig:pfasst} 
\end{figure}

Both approaches are depicted in Figure~\ref{fig:pfasst} and it becomes obvious that in terms of performance, the classical view is advantageous, since idle times of the processes are largely avoided.
The code, however, is more complex to write and extend, while in principle the multigrid view allows PFASST to be written using any adequately flexible multigrid framework.
Also, it should be noted that it is much more straightforward and shorter to derive PFASST as an FAS multigrid and it also allows to analyze PFASST's properties much more easily, see e.g.~\cite{BoltenEtAl2017}.

\section{The pySDC framework}

We have seen in the previous section, how SDC and PFASST can be derived, how they work and how their mathematical descriptions look like.
For both algorithms, there are a multitude of algorithmic choices to make and many parameters to define.
For SDC alone, not even considering the spatial problem, we need to select the time-step size, specify type and number of collocation nodes, the preconditioner $\QDmat$ and the number of iterations or the threshold for the residual.
For PFASST, at least the number of time-steps, the number of levels, the coarsening strategy and restriction and prolongation need to be selected.
Then, the spatial problem defining the system of ODEs has to be specified in a way that SDC and PFASST can work with it.
More precisely, we need to be able to evaluate the right-hand side $f$ of the ODE system and, if parts of $f$ should be treated implicitly, an implicit solver has to be provided, including all the parameters these choices may introduce.
A code which implements SDC, PFASST or any other solver of collocation problems needs to provide ways to account for this multitude of choices, giving the users as much flexibility as possible while exposing only as few internals as absolutely necessary.

We will describe in the following key components and features of \pysdc{}, starting with the overall structure of the code and its documentation. We then take a user's perspective and show a simple yet typical code snippet a user might write for her or his application. The last part takes the developer's perspective, describing the main components of \pysdc{} and how the user script is used to set up and run SDC or PFASST variants.

\subsection{Availability and structure}

Starting with version 2, the main website for \pysdc{} can be found under
\begin{center}
\url{http://www.parallel-in-time.org/pySDC}.
\end{center}
Here, all relevant information can be obtained, including links to the code repository on Github~\cite{pysdc_github} and the Python Package Index (PyPI) version~\cite{pysdc_pypi}, the documentation as well as test coverage reports.
\pysdc{} requires some Python packages to be installed, especially NumPy~\cite{numpy}, SciPy~\cite{scipy} and Matplotlib~\cite{matplotlib} are used throughout the package. 
All dependencies are listed in the \texttt{requirements.txt} file.
In particular when using PyPI and \texttt{pip}, installation of \pysdc{} should work without any problems. 

The documentation contains two parts: (1) the API documentation generated automatically by Sphinx and (2) the documentation of the projects and the tutorial. 
While (1) is created using the Python comments, the text of (2) is written separately for each project and each tutorial and contains codes and results. 
Both API and example documentation is generated during automated testing after each update of the master branch using a Github hook to Travis CI~\cite{travis}.
This continuous integration engine runs the test suite, creates the documentation, updates the website and the PyPI version, if necessary.
In addition, PEP8 conformity is guaranteed and tested with a few minor exceptions.
The code is tested to work with Python 2.7 and 3.6 under Ubuntu Trusty (as of August 2018) and test coverage reports are shown online. 
It is reported to work on recent macOS installations with various Python versions as well.

The \pysdc{} package is divided into seven subfolders containing various components of the framework. In detail:

\begin{itemize}
\item \texttt{core}: this folder contains the core components of \pysdc{}, providing the basis for deriving custom functionality such as sweepers, problem classes, and transfer operators, but also the main structures \texttt{level} and \texttt{step}.
\item \texttt{implementations}: in various subfolders, this part contains all implementations of \pysdc{}'s functionality. For example, problem classes, data types, collocation classes are defined and provided for further use. This folder should contain all implementations of relevance for multiple examples or of general interest.
\item \texttt{projects}: this folder consists of all larger projects leading to publications or suitable for further demonstration. There are also specialized versions of the implementations which are used exclusively for the purpose of a single project.
\item \texttt{playgrounds}: the playgrounds are a loose collection of smaller projects and code snippets. In contrast to the projects, this part is not necessarily documented and PEP8 conformity is not tested.
\item \texttt{tutorial}: here, the codes for seven tutorials are located to provide an easier access to the theory, the code, and its functionality. The tutorials range from simple SDC runs to multi-level setups and MPI-parallel PFASST studies. Each tutorial has a specified outcome and the whole set of introductory codes is tested each time before deployment. 
\item \texttt{tests}: this folder contains the tests to be run for the whole package. These consist of simple executions of tutorials and parts of the projects as well as more direct tests of the core functionality of \pysdc{}.
\item \texttt{helpers}: here, smaller tools and scripts are located, e.g.\ helping users to plot or evaluate statistics.
\end{itemize}

Not all parts are relevant for all types of users. 
In the following we therefore shed light on the code from two different perspectives.

\subsection{A user's perspective}

A good starting point for users are \pysdc{}'s seven tutorials.
While the auto-generated API documentation formally describes the main functions, the tutorials show how to actually work with them. 
We will not go into details here, since all information can be also found online. 
Instead, we discuss the fundamental structure of a \pysdc{} application script. 
Passing user information to the components of \pysdc{}, e.g.~to sweepers or the controller, but also to the user-defined problem class is implemented using Python dictionaries. 
Each component expects a dictionary during instantiation, where certain parameters have to be present. 
Many parameters already have pre-defined values and new parameters can be added by the user.

In Listing~\ref{lst:pysdc_main} a typical "main" routine for applications is shown.
This listing would execute serialized two-level PFASST for the one-dimensional heat equation with forcing term and a certain choice of parameters. 
The first 23 lines simply create and fill dictionaries for the problem, the sweeper, the level, the sweeper, the step and the controller.
In lines 26-36, the main dictionary called \texttt{description} is then filled with the parameter dictionaries but also with classes (and not instances). 
Note that these classes as well as the controller class used below have to be imported e.g.\ from the \texttt{implementations} resources.
This dictionary together with the controller parameters fully define the setup of \pysdc{} for a particular problem.
In line 39, then, the controller is instantiated, where also the number of parallel steps (\texttt{num\_procs} in this context) is set.
Providing some initial values in line 42, which have to be of type \texttt{dtype\_u}, see line 29, the code is run in line 45, where also start and end times are given.
The controller returns the values at the final time as well as a \texttt{stats} dictionary, which contains a lot of information about the run, e.g.\ intermediate residuals, timings and so on.
This dictionary can be analyzed using tools from the \texttt{helper} folder. 
Examples are shown in the tutorial and the projects.

Note that we use the so called \texttt{allinclusive\_multigrid\_nonMPI} controller here, which 
\begin{enumerate}
\item can do SDC, MLSDC and PFASST just by using different parameters ("allinclusive"),
\item is written in a multigrid-fashion, following the more recent interpretation of PFASST ("multigrid"),
\item and is not parallel, i.e.\ the code is executed in serial but works like a parallelized code ("nonMPI").
\end{enumerate}
The first attribute is explained in more detail below, the second is a mere technical and historical terminology (the "classical" controllers, following the original description of PFASST, have become deprecated in a previous release of \pysdc{}), and the third attribute makes is much easier to test, debug and study the algorithm. 
MPI-parallel versions ca be found in the \texttt{implementations} folder, too.

\lstset{language=Python}
\lstset{frame=lines}
\lstset{caption={Main routine for applications}}
\lstset{label={lst:pysdc_main}}
\begin{lstlisting}
# initialize problem parameters
problem_params = dict()
problem_params['nu'] = 0.1		# diffusion coefficient
problem_params['freq'] = 8		# frequency for the test value
problem_params['nvars'] = [511, 255]    # number of degrees of freedom 

# initialize sweeper parameters
sweeper_params = dict()
sweeper_params['collocation_class'] = CollGaussRadau_Right
sweeper_params['num_nodes'] = 3

# initialize level parameters
level_params = dict()
level_params['restol'] = 1E-10
level_params['dt'] = 0.25

# initialize step parameters
step_params = dict()
step_params['maxiter'] = 50

# initialize controller parameters
controller_params = dict()
controller_params['logger_level'] = 20

# fill description dictionary for easy step instantiation
description = dict()
description['problem_class'] = heat1d_forced	# pass problem class
description['problem_params'] = problem_params  # pass problem parameters
description['dtype_u'] = mesh 			# pass data type for u
description['dtype_f'] = rhs_imex_mesh 		# pass data type for f
description['sweeper_class'] = imex_1st_order 	# pass sweeper
description['sweeper_params'] = sweeper_params  # pass sweeper parameters
description['level_params'] = level_params      # pass level parameters
description['step_params'] = step_params        # pass step parameters
description['space_transfer_class'] = mesh_to_mesh  # pass spatial transfer class
description['space_transfer_params'] = dict() 	# no params for spatial transfer 

# instantiate controller
controller = allinclusive_multigrid_nonMPI(num_procs=8, controller_params=controller_params, description=description)

# set initial values on finest level
uinit = ...

# call main function to get things done...
uend, stats = controller.run(u0=uinit, t0=0.0, Tend=4.0)
\end{lstlisting}

If a user wants to work with an already existing problem class, this script is the only part where modifications are required.
Of course, \pysdc{} allows to define custom problem classes, data types, transfer operators and so on.
Imported in such a main routine, they can be passed to \pysdc{} in a very simple manner.
In particular, the user is free to use her or his own data types and spatial solvers, which do not even have to be written in Python.
\pysdc{} only requires basic operations with data types, from creating new instances to addition, multiplication by a scalar and some sort of absolute value.
\pysdc{} already comes with \texttt{mesh} and \texttt{particle} data types and also  couplings to the FEniCS framework via DOLPHIN~\cite{Logg2010-em,Logg2012-lo} as well as to petsc4py~\cite{DALCIN20111124} have already been implemented.
Again, examples can be found in the tutorials and the projects.

There is one particular aspect of the parameter handling we want to highlight here, though.
In line 5, a list of numbers is given as the degrees-of-freedom for the problem class. 
This is the way to indicate the usage of multiple levels, two in this case. 
Whenever the controller instantiation finds a list of objects within a parameter dictionary of the \texttt{description}, it will create multiple levels for the run.
One could for example specify \texttt{[5,3,2]} collocation nodes instead of just \texttt{3} in line 10. 
Then, the controller would create three levels with $5$, $3$ and $2$ nodes and $511$, $255$ and $255$ degrees-of-freedom.
This also applies to lists of problem classes, so that in line 27 a list of different problem classes could be provided, having e.g.\ different spatial solvers on the different levels.
This way the main routine and the controller do not change much when going from SDC with a single level to MLSDC with multiple levels and PFASST with multiple levels and time-steps (and this is why the controller has the attribute "allinclusive"). 
To decide whether to run PFASST or MLSDC only the \texttt{num\_procs} parameter, see line 39, has to be changed.

\subsection{A developer's perspective}

\subsubsection{Setup}\label{sssec:setup}

When the code in Listing~\ref{lst:pysdc_main} is run with the current logging level, the output starts with an overview of all parameters (default and user-defined) as well as the hierarchy of levels.
This tells the user, which parameters are actually available and which were provided during initialization.
The output however also provides an overview of the structure of the controller and all underlying components. 
For the example above the structure looks like this ("..." here indicates long import paths, mostly from the \texttt{implementations} folder):
\begin{description}
  \item[Controller] \texttt{class '...allinclusive\_multigrid\_nonMPI'}
  \begin{description}
    \item[Step] \texttt{class 'pySDC.core.Step.step'}
    \begin{description}
      \item[Level 0] \texttt{class 'pySDC.core.Level.level'}
        \begin{description}
          \item[Problem] \texttt{class '...heat1d\_forced'}
          \begin{description}
            \item[Data type solution] \texttt{class '...mesh'}
            \item[Data type RHS] \texttt{class '...rhs\_imex\_mesh'}
          \end{description}
          \item[Sweeper] \texttt{class '...imex\_1st\_order'}
          \begin{description}
            \item[Collocation] \texttt{class '...CollGaussRadau\_Right}
          \end{description}
        \end{description}
        \item[Level 1] \texttt{class 'pySDC.core.Level.level'}
        \begin{description}
          \item[Problem] \texttt{class '...heat1d\_forced'}
          \begin{description}
            \item[Data type solution] \texttt{class '...mesh'}
            \item[Data type RHS] \texttt{class '...rhs\_imex\_mesh'}
          \end{description}
          \item[Sweeper] \texttt{class '...imex\_1st\_order'}
          \begin{description}
            \item[Collocation] \texttt{class '...CollGaussRadau\_Right}
          \end{description}
        \end{description}
        \item[Base transfer] \texttt{class 'pySDC.core.BaseTransfer.base\_transfer'}
        \begin{description}
          \item[Space transfer] \texttt{class '...mesh\_to\_mesh'}
        \end{description}
    \end{description}
  \end{description}
\end{description}

Upon initialization, the controller creates \texttt{num\_procs} steps. 
Each step then creates the correct number of levels derived from the user's input values.
Within each level, a problem and a sweeper is created, where the problem contains the actual data types and the sweeper the collocation class.
The step initialization then connects two consecutive levels using the base transfer class, which in turn has a space transfer instance as a member.

Each of these components can be changed. 
While problem, sweeper, collocation, data type and spatial transfer classes are provided by the user within the \texttt{description} dictionary, changing the core modules require deeper intervention.
Nevertheless, if for example the transfer class (where among others the $\tau$-correction of the FAS is computed) should behave differently, this part can be exchanged independently and rather easily.

Besides this hierarchy, the controller also creates an instance of a so called hook class as well as a logger object. While the latter can be used to take care of file and terminal status output, the first provides initially some basic means to add statistics during the run.
Hook methods are called e.g.\ before and after each step, each iteration, and each run. 
In the original hook class in the \texttt{core} folder, only basic information on the current status is added to the \texttt{stats} dictionary the controller returns after the run.
However, this class can be overridden by the user or developer to add much more functionality like visualization, steering, or more advanced statistic like counting the number of iterations of a spatial solver across multiple calls.

\subsubsection{Execution}

After the hierarchy of objects, the logger and the hooks are created, the controller has access to all functionality and can use methods of all classes at will.
Ideally, however, only methods on the sweeper level should be accessed by the controller, in contrast to methods of the problem class or even data types.
This design principle is applied to all levels of \pysdc{}.
For example, the sweeper does not need to have access to problem class details, apart from calling right-hand side evaluations and solvers, if any. 
This can be seen in Eq.~\eqref{eq:sdc_iteration_node}, where some generic right-hand side $f$ as well as a solver for the implicit part, provided that $\QDmat$ has entries on the diagonal, are needed.
In addition, the sweeper only needs to know how data types are added or scaled, but not which data it actually handles.
In the very same manner, the controller only calls sweeper routines to update nodes (which is the actual sweep) or to build the residual, but how this is actually implemented or what kind of sweeper is used is not relevant for the controller.
If time-parallelism is requested, communication routines have to be accessible, too, which are part of the data types.

After setting up the infrastructure during the initialization of the controller, the \texttt{run} routine can be called.
This is a single procedure for running as many steps of SDC, MLSDC or PFASST with all the particular parameters as requested by the user.
It starts by assigning each of the \texttt{num\_procs} steps a particular time-step, including a start time and a time-step size $\dt$.
Different $\dt$ are allowed throughout the controller to enable adaptivity in time, although actual adaptive algorithms are not yet part of \pysdc{}.
The controller checks which step is active, i.e.\ which time-step is still within the interval $[0,T]$.
Then, for these active steps the main time loop is entered, where each time-step in the case of single-step methods like SDC or blocks of time-steps in the case of multi-step methods like PFASST is treated.
After convergence has been declared, e.g.\ by reaching a particular tolerance or the maximum number of iterations, the start times and time-step sizes are updated, active steps are identified and the next round of the loop is entered, if there are any active steps left.
Otherwise, the controller exits and returns the final value as well as the \texttt{stats} dictionary.

The inner loop, i.e.\ the treatment of a single time-step or of a block of time-steps, is done in so called "stages".
This is implemented by splitting the algorithm into different parts (the stages), where each part ends by defining what to do next.
As long as the stages of all active time-steps are not set to "done", the controller keeps working on them.
This implementation is clearly not the most efficient one, but we have identified three major advantages:
\begin{enumerate}
	\item flexibility, since the sequence of the stages can be changed or extended rather easily,
    \item readability, since the quite complex algorithm is split into reasonable units,
    \item maintainability, since the current stage can be accessed or shown at any time.
\end{enumerate}

\begin{figure}
	\includegraphics[width=\textwidth]{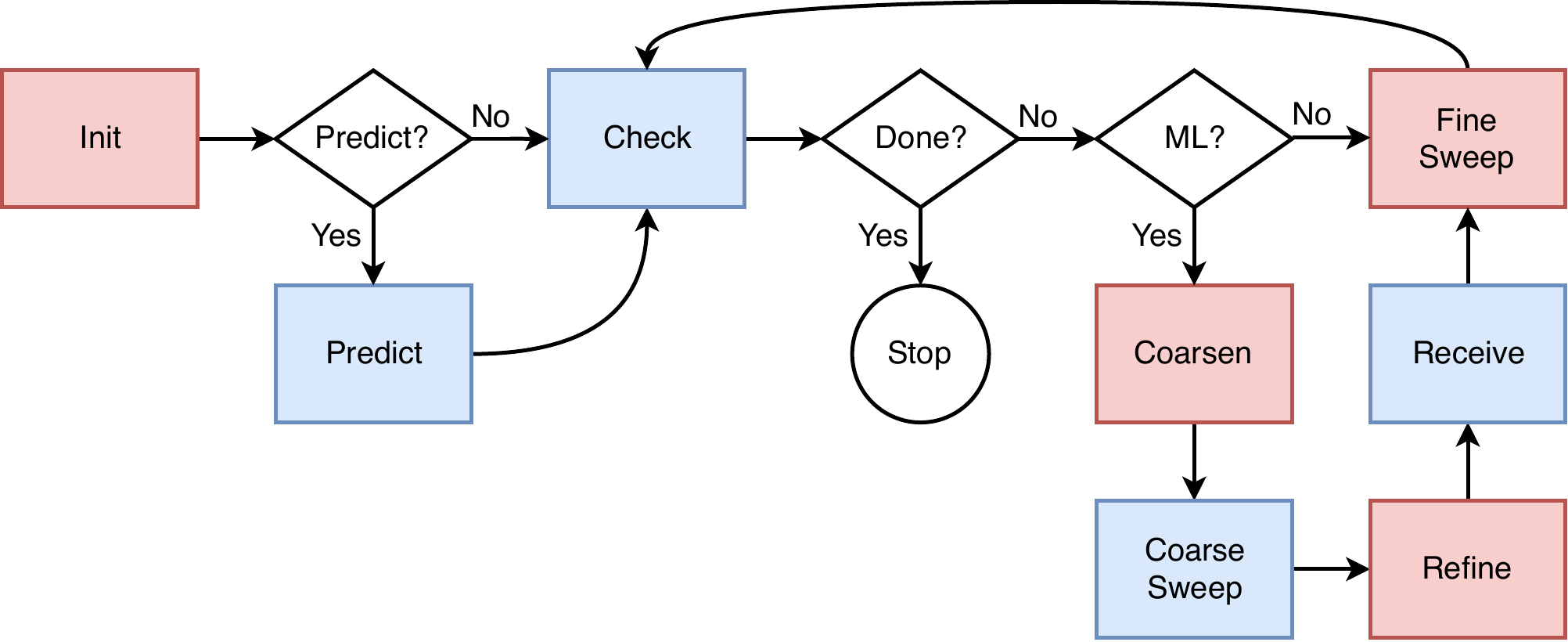}
    \caption{Basic workflow of the allinclusive controllers. Red/light color: can be executed in parallel, blue/dark color: contains blocking communication if PFASST is run. "ML?" asks for multiple levels. "Stop" refers to the current time-step or block of time-steps.}
    \label{fig:workflow}
\end{figure}

Figure~\ref{fig:workflow} gives a (somewhat simplified) overview of the different stages.
In blue/dark colors, stages with potentially blocking communication are indicated.
Stages in red/light colors can be computed fully in parallel, maybe including non-blocking forward communication of data.
All active time-steps start with some initialization phase and proceed from there.
A prediction routine can be specified as part of the current controller and then the main loop is entered by checking for convergence first.
The controller can run on a single or on multiple levels, depending on the input parameters.
We note that in principle the controller could run on a single level but with multiple steps (i.e.\ "single-level PFASST" or "multi-step SDC"), but this has not been implemented yet.

If developers would like to change this workflow, a new controller class can be derived from any of the existing controllers.
Since the definition of the stages including their order is defined in a separate routine, changing the order, changing existing stages or adding new ones can be done without having to deal with the outer control logic.
Of course, overriding this outer logic is also possible, e.g.\ for introducing adaptivity or adding workflow-specific features.
One example where this has been done is the fault-tolerant PFASST controller, which is part of the \texttt{node\_failure} project and which will be discussed in Sect.~\ref{ssec:ex_ft}.

\section{Numerical experiments}

In this section we will present three examples where \pysdc{} was used to run SDC and PFASST simulations.
Each example has a different focus and highlights different aspects of the usage of \pysdc{}.

\subsection{SDC for the Allen-Cahn equation}

In the first example we compare different variants of SDC for the two-dimensional Allen-Cahn (or reaction-diffusion) equation
\begin{align}\label{eq:ac}
	u_t &= \Delta u + \frac{1}{\epsilon^2} u(1-u^2)\quad \mathrm{on}\quad [-0.5, 0.5]^2\times[0,T],\ T>0,\\
    u(x,0) &= u_0(x),\quad x\in[-0.5,0.5]^2,\nonumber
\end{align}
with periodic boundary conditions and scaling parameter $\epsilon > 0$.
Following~\cite{zhang_SISC_AC}, we choose 
\begin{align*}
	u_0(x) = \mathrm{tanh}\left(\frac{R_0 - \lvert x\rvert}{\sqrt{2}\epsilon}\right),
\end{align*}
which describes a circle with initial radius $R_0 > 0$ and interface width $c\epsilon$ for some constant $c>0$
This profile is instable and the circle will shrink over time.
In the sharp interface limit $\epsilon \rightarrow 0$ the radius can be expressed as
\begin{align*}
	r(t) = \sqrt{R_0^2 - 2t}
\end{align*}
and thus provides a very good indicator whether or not the numerical scheme captures the dynamics of this equation correctly.

We use simple second-order finite differences for discretizing in space with $N=128$ and $\epsilon=0.04$, so that initially about $5$ points resolve the interface, which has a width of about $7\epsilon$.
We furthermore use $M=3$ Gauss-Radau nodes and $\dt=0.001 < \epsilon^2$ for the collocation problem and stop the simulation after 32 time-steps at $T=0.032$, since then $r(t) = 0$ for $R_0=0.25$.

Five different variants of SDC are tested with \pysdc{}:
\begin{enumerate}
	\item fully-implicit SDC, where the full right-hand side of~\eqref{eq:ac} is treated implicitly,
    \item semi-implicit SDC, where the diffusion term is treated implicitly and the rest explicitly
    \item multi-implicit SDC, where both reaction and diffusion term are treated implicitly but decoupled, see~\cite{BOURLIOUX2003651} for details,
    \item modified semi-implicit SDC, where the dissipative term $\Delta u - \epsilon^{-2}u^3$ is treated implicitly and the expansive term $\epsilon^{-2}u$ explicitly, see~\cite{GraeserEtAl2013} for details,
    \item modified multi-implicit SDC, where both dissipative and expansive terms are treated implicitly but in a decoupled manner.
\end{enumerate}
All variants use the LU-trick for the implicit part(s) and the explicit Euler for the explicit ones.
We stop the iteration when a residual tolerance of $10^{-8}$ is reached.
Moreover, each variant is tested with (a) full and (b) reduced accuracy in the linear and/or nonlinear implicit solvers. 
The latter approach is called inexact SDC and has been investigated in~\cite{SpeckEtAl2016,WeiserGhosh2018}.
We use Newton's method for the nonlinear systems and CG for the linear systems. 
Exact variants iterate until a tolerance of $10^{-9}$ for Newton and $10^{-10}$ for CG is reached. 
Inexact variants stop after $1$ Newton and $10$ CG iterations.

\begin{figure}[t]
  \begin{subfigure}[b]{0.49\textwidth}
    \centering
    \includegraphics[width=0.95\columnwidth]{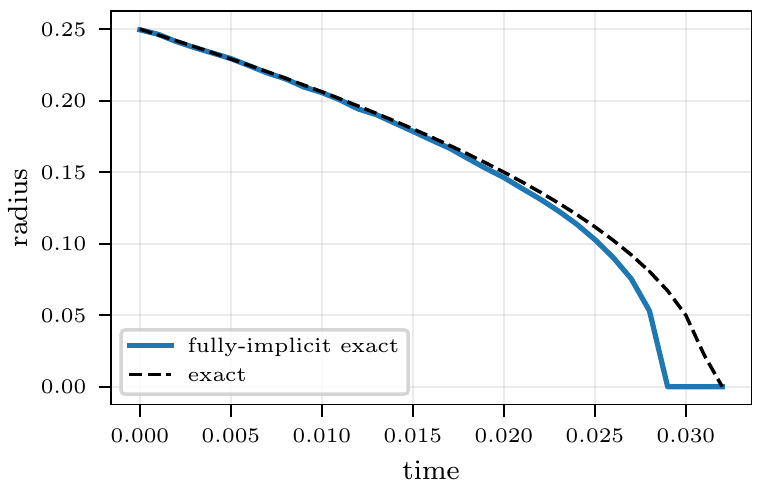}
    \caption{Circle radius over time.}
    \label{fig:ac_radii}
  \end{subfigure}
  \hfill
  \begin{subfigure}[b]{0.49\textwidth}
    \centering
    \includegraphics[width=0.95\columnwidth]{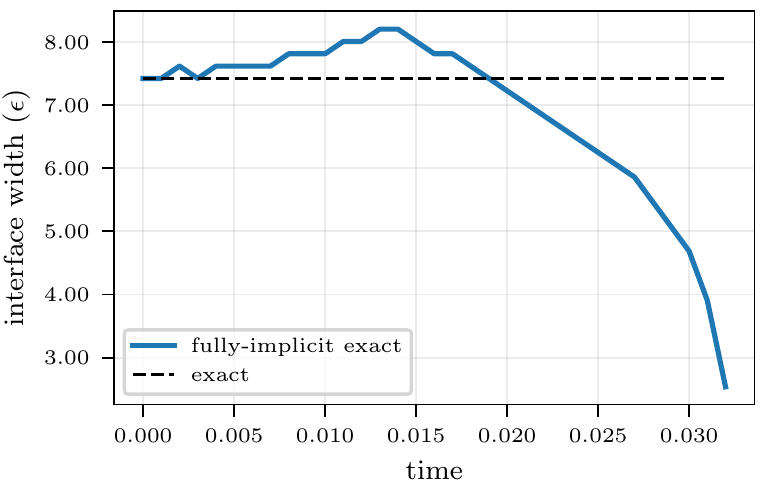}
    \caption{Interface width over time.}
    \label{fig:ac_interface}
  \end{subfigure}
  \caption{Dynamics of the Allen-Cahn equation. All other SDC variants show exactly the same behavior.}
  \label{fig:ac_dynamics} 
\end{figure}

Before we look at the timings and iteration counts, we check in Figures~\ref{fig:ac_radii} and~\ref{fig:ac_interface} the evolution of the radius and the interface width over time.
Until about $t=0.02$, the computed radius is very close to the exact one, see Figure~\ref{fig:ac_radii}.
After that, the computed radius goes to zeros slightly faster, i.e. the circle vanishes faster than it should. 
The reason for that can be found in Figure~\ref{fig:ac_interface}: the computed interface width remains above $7\epsilon$ until $t\approx0.02$, but after that the interface is not resolved sufficiently anymore due to the rather simplistic spatial discretization.
A more advanced method as e.g. in~\cite{zhang_SISC_AC} would provide a better result in this case.
Nevertheless, the code produces meaningful results and more notably, all variants produce exactly the same curves (not shown here).
Thus, all 10 variants are able to capture the same dynamics, which, at least until $t\approx0.02$, is very close to the correct one.
In particular, even seemingly unfavorable SDC variants like standard semi-implicit splitting (see the discussion in~\cite{GraeserEtAl2013}) yield the correct results, since they all converge to the solution of the fully-implicit collocation problem.

\begin{figure}[t]
  \centering
  \includegraphics[width=0.95\columnwidth]{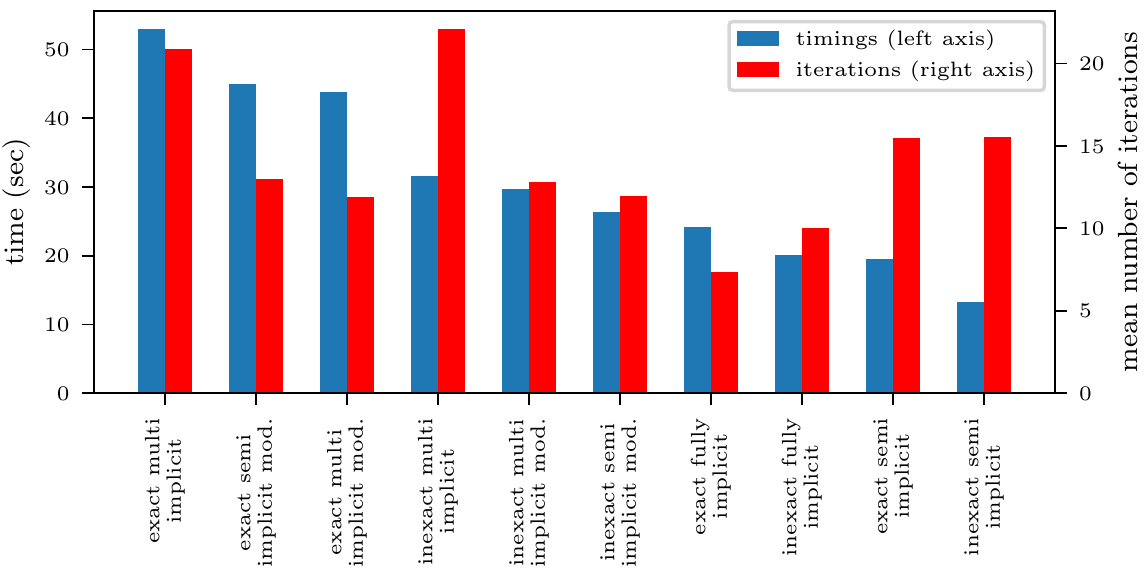}
  \caption{Timings and mean number of iterations for different SDC variants solving the Allen-Cahn equation.}
  \label{fig:ac_timings}
\end{figure}

In Figure~\ref{fig:ac_timings} we show the timings (blue/left bars, left y-axis) and iteration counts (red/right bars, right y-axis) for all variants of SDC. 
We can see a factor of about 5 between the slowest variant ("exact multi-implicit") and the fastest one ("inexact semi-implicit").
We also see that this does not correspond to the number of iterations but rather to the cost of those iterations. 
For example, the inexact semi-implicit variant uses only 10 CG iterations per SDC iteration, but requires about twice as many SDC iterations as the exact fully-implicit variant, which in turn takes twice as long due to expensive Newton iterations.

We note that of course many further ideas could be exploited in this context, ranging from parallel SDC, MLSDC, and PFASST to more advanced spatial solvers~\cite{GraeserEtAl2009} or adaptive selection of tolerances for inexact SDC~\cite{WeiserGhosh2018}. 
However, for this paper and for \pysdc{} the more relevant aspects are to show that implementing, testing and running multiple variants of SDC (and MLSDC and PFASST for that matter) is straightforward.
Also, this example highlights the potential of computing and gathering statistics such as radii or interface width on the fly using hooks, see the discussion at the end of Section~\ref{sssec:setup}.

The code for this example can be found in the folder \texttt{projects/TOMS} of \pysdc{}~\cite{pySDC_release}.

\subsection{Fault-tolerant PFASST with FEniCS}\label{ssec:ex_ft}

In the second example we study algorithm-based fault tolerance of PFASST, in particular recovery after processor failure (so called "hard faults").
To this end we use \pysdc{} to emulate the crash of a processor which leads to the loss of data at one particular time-step.
We assume immediate replacement of the processor and then restart the iteration for the affected time-step.
This example and most of the details given below are taken from~\cite{SPECK201720}.

We consider the 1D Gray-Scott model~\cite{Gray1983-im} for a chemical reaction of two components $\mathcal{U}$ and $\mathcal{V}$ with
\begin{equation}
    \label{eq:grayscott}
    \begin{aligned}
        u_t &= \Delta u - uv^2 + A(1-u),\\
        v_t &= D\Delta v + uv^2 - Bu,
    \end{aligned}
\end{equation}
where $u = u(x,t)$ and $v = v(x,t)$ are the concentrations of the two species $\mathcal{U}$ and $\mathcal{V}$, $D$ is the normalized diffusion coefficient of $\mathcal{V}$, $A$ denotes the fed rate into the system (e.g.~a reactor) and $B$ is the overall decay rate of $\mathcal{V}$.

For our simulations, we choose $A=0.09$, $B=0.086$, $D=0.01$, and start with initial conditions
\begin{equation}
    u(x,0) = 1 - \frac{1}{2}\sin^{100}(\pi x/L),\quad v(x,0) = \frac{1}{4}\sin^{100}(\pi x/L)
\end{equation}
for $x\in[0,L]=[0,100]$ in our case, representing a sharp initial peak at the center of the domain, see~\cite{Doelman1997-zu}. 
We use homogeneous Neumann boundary conditions.

For the spatial discretization, we make use of the FEniCS framework~\cite{Logg2012-lo} and in particular the interface DOLFIN~\cite{Logg2010-em} with its Python front-end.
\pysdc{} allows to handle FEniCS' weak formulation of PDEs and is thus capable of handling complicated multi-component equations by exploiting FEniCS' formalism.
By specifying the right-hand sides in weak form and using FEniCS's built-in solvers, \pysdc{} provides easy-to-use high-order time-stepping for finite element discretizations.

We choose $3$ Gauss-Radau collocation nodes and the simulation is run until to $T=1280.0$ with $\Delta t = 2.0$, i.e.~for $640$ time steps.
We parallelize in time using $20$ blocks of $32$ parallel steps with two-level PFASST. 
Each block iterates until a residual of $10^{-7}$ is reached (absolute tolerance).
We use fourth-order standard finite elements with $N_f=513$ degrees-of-freedom on the fine and $N_c=257$ on the coarse level.
FEniCS' built-in Newton's method serves as spatial implicit solver with absolute tolerance $10^{-9}$ and relative tolerance $10^{-8}$, treating the full right-hand side of the PDE implicitly, using again the LU-trick for $\QDmat$.

\begin{figure}[t]
  \begin{subfigure}[b]{\textwidth}
    \centering
    \includegraphics[width=\columnwidth]{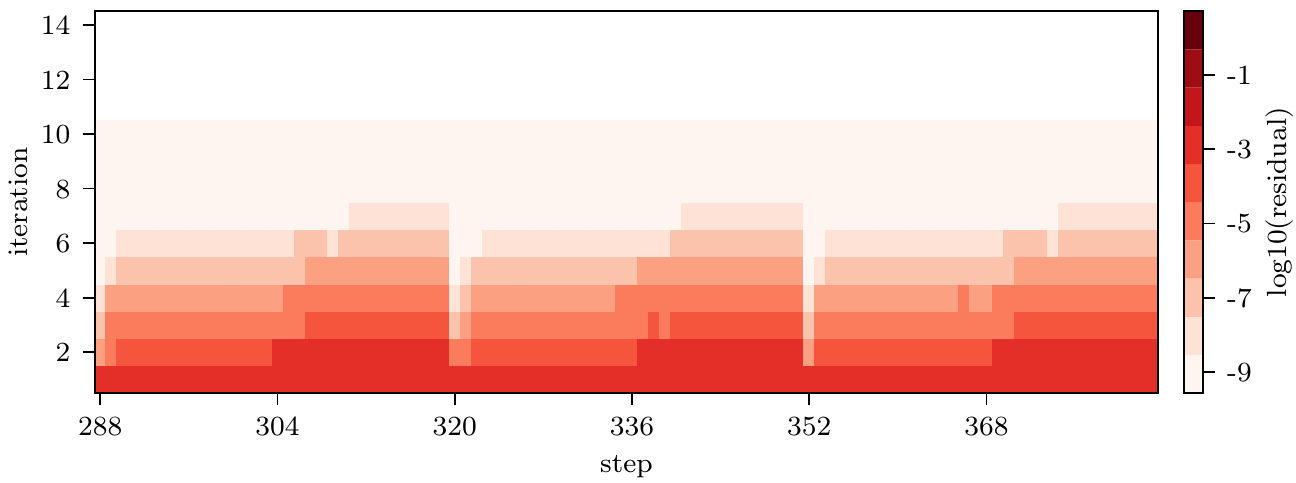}
    \caption{Gray-Scott run without faults.}
    \label{fig:gs_nofault}
  \end{subfigure}
  \par\bigskip
  \begin{subfigure}[b]{\textwidth}
    \centering
    \includegraphics[width=\columnwidth]{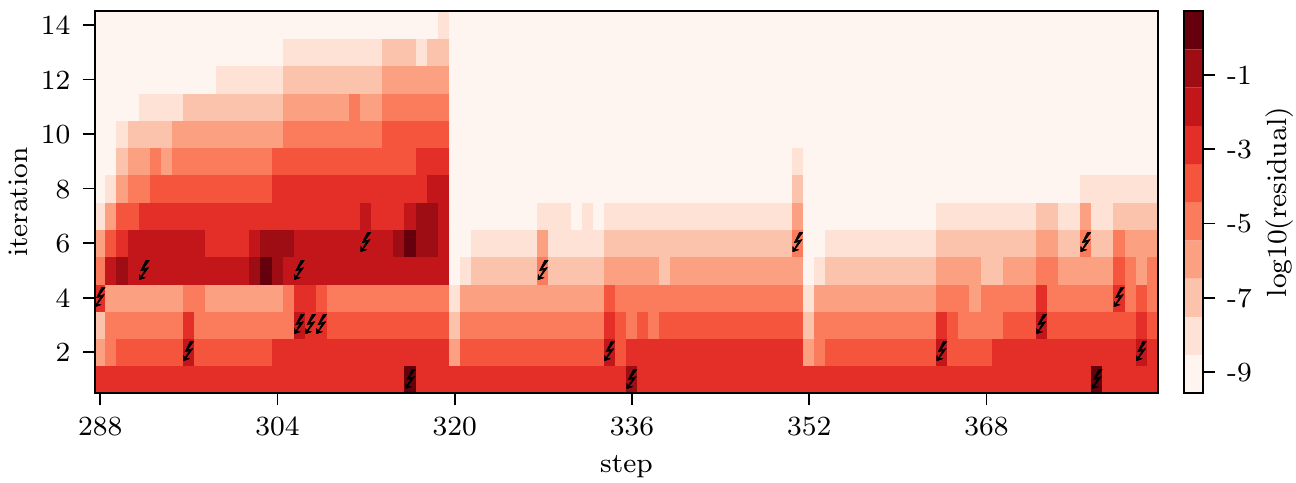}
    \caption{Gray-Scott run with faults and interpolation restart.}
    \label{fig:gs_interp}
  \end{subfigure}
  \caption{Residuals for three blocks of PFASST with 32 parallel time-steps for the Gray-Scott equation. Fault-free scenario on top and recovery using two-sided interpolation to restart below.}
  \label{fig:gs_ft} 
\end{figure}

For the no-fault run, the maximum number of iterations is $K=7$ except for the first block which needs $K=10$ iterations due to its fast dynamics. 
In Figure~\ref{fig:gs_nofault} we show the residuals of this fault-free run for three representative time-parallel blocks, ranging from $t=578$ to $t=768$ (blocks $9$, $10$, $11$) which contain time steps $288$ to $384$.
To ``stress-test'' PFASST, faults are then injected at random. 
Before starting a new iteration, we inject a fault with a (hopefully unrealistically high) probability of $3\%$.
The random procedure creates different failure patterns throughout the run, which are difficult to anticipate and create a-priori.
Moreover, it minimizes the chance of testing the recovery strategies for favorable conditions only.
Interesting structures observed in the distribution of faults are e.g.~multiple faults of the same process in quick succession, which, in a real-world system, could be caused by a faulty component or clusters of faults mimicking cascading failures.
In order to be able to compare different strategies, the random pattern of faults is generated a-priori and then applied to all simulations testing different recovery strategies.
To this end we created a new controller within the \pysdc{} project which allows to inject faults and to restart a time-step afterwards.
To order to avoid dealing with the actual restart procedure (detection, fault-tolerant MPI, assignment of a new processor, initial data transfer etc.) and to focus on the methodological aspects, we use the serial implementation of PFASST (the "nonMPI" variant) as basis, which has access to all data at all time-steps.
We only allow faults in iterations that are also performed in the no-fault case.

To recover after a fault, the data at collocation nodes of the affected time-step is restored by interpolating between the data from the previous and the next time-step.
This already works rather well, but the impact can be increased by adding correction sweeps on the coarse level. 
For details on this and other recovery strategies we refer to~\cite{SPECK201720}.
The result of this algorithm-based recovery strategy can be seen in Figure~\ref{fig:gs_interp}.
Faults are indicated by the little lightning symbols.
In the first block shown (steps $288$--$320$), up to $K_\text{add}=7$ additional iterations are required to converge.
This is the worst case throughout the whole run and due to the cluster of failures in iterations $3$ and $5$. 
While the two isolated faults in iterations $1$ and $2$ have negligible impact on convergence, the three failures in iteration $3$ as well as the failures in iterations $5$ and $6$ lead to very high residuals on all subsequent processes.
In the next block, only four faults are injected and their impact is limited, leading to $K_\text{add}=2$ additional iterations which are mainly due to the isolated faults in iterations $5$ and $6$.
The last block shows a cluster of faults: during six iterations there are four faults within the last seven processes. 
However, only a single additional iteration is needed to finish, underlining the effectiveness of this recovery strategy.

Again, this study of algorithm-based fault tolerance with \pysdc{} leaves a lot of room for improvement as well as many open questions (some of which are already stated in~\cite{SPECK201720}). 
In particular, by using \pysdc{} and one of its serial PFASST controllers we completely avoided the hardships of actual MPI-parallel fault-tolerant implementations. 
However, this is exactly the point of \pysdc{} and this example: a prototyping framework allows users to test their ideas without the need to invest a lot of time thinking about a (potentially useless) implementation.

The code to reproduce the results of this part can be found in the \texttt{projects/hard\_fault} folder of \pysdc{}~\cite{pySDC_release}.

\subsection{Space-time parallel PFASST with PETSc}

The goal of the last example is to show, at least to a certain degree, that users of \pysdc{} can go beyond prototyping their algorithms and are able to set up actual space-time parallel runs without the need of lower-level programming languages like Fortran or C++.
Besides the FEniCS framework, \pysdc{} also comes with a fully-functional link to PETSc~\cite{petsc-user-ref} through petsc4py~\cite{DALCIN20111124}, which includes access to PETSc data types, solvers and MPI-based parallelization in space.

To demonstrate this combination of parallelism in space and time, we resort to the standard test case for parallel-in-time integration methods~\cite{Gander2015_Review}, i.e. the forced heat equation in 2D with
\begin{align*}
	u_t &= \Delta u + f\quad \mathrm{on}\quad [0, 1]^2\times[0,T],\ T>0,
\end{align*}
with forcing term $f(x,t)$ such that the exact solution is given by
\begin{align*}
	u(x,y,t) = \sin(2\pi x)\sin(2\pi y)\cos(t)\quad\mathrm{on}\quad [0, 1]^2\times[0,T]
\end{align*}
and the initial condition is given simply by $u(x,y,0)$.
We use homogeneous Dirichlet boundary conditions.

Standard finite differences of second-order are used for the spatial discretization with $N=129$ degrees-of-freedom on the fine and $65$ on the coarse level, $M=5$ Gauss-Radau nodes in time and $\dt=0.125$ and $T=3.0$.
We apply semi-implicit SDC as the base methods, treating the diffusion term implicitly using the LU trick and the forcing term explicitly.
The tolerance for the outer MLSDC and PFASST residual is set to $10^{-8}$ for all runs, the inner linear solver stops at a residual below $10^{-12}$.

With PETSc as the backend, we can now use DMDA Global Vectors as data type instead of a Numpy array and the linear KSP solvers instead of SciPy ones.
Therefore, distributed arrays across multiple MPI ranks and scalable solvers are available, making space-time parallelism possible by splitting the MPI\_COMM\_WORLD communicator accordingly.
This allows to extend the prototyping approach of \pysdc{} from a more method-oriented idea to prototyping and benchmarking of space-time parallelization strategies or controller implementations.

\begin{figure}[t]
  \begin{subfigure}[b]{0.49\textwidth}
    \centering
    \includegraphics[width=0.9\columnwidth]{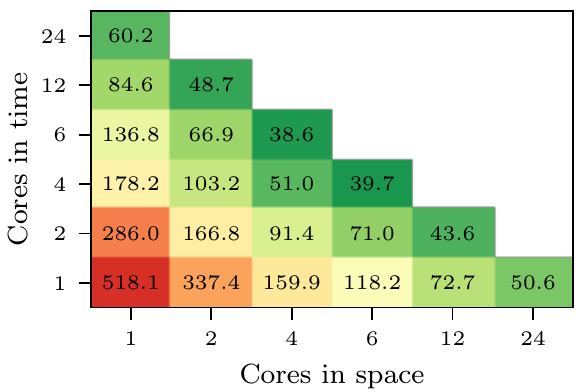}
    \caption{Timings for various distributions of cores.}
    \label{fig:petsc_matrix}
  \end{subfigure}
  \begin{subfigure}[b]{0.49\textwidth}
    \centering
    \includegraphics[width=0.9\columnwidth]{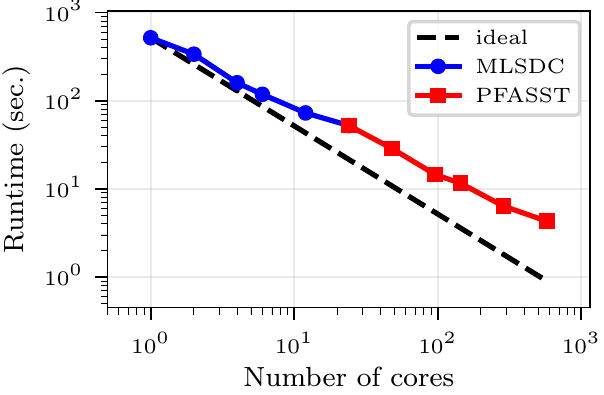}
    \caption{Speedup of time- over space-parallelism.}
    \label{fig:petsc_speedup}
  \end{subfigure}
  \caption{Benchmarks of the heat equation on JURECA using PETSc for spatial parallelization.}
  \label{fig:petsc} 
\end{figure}

To demonstrate this, we run the heat equation example on the JURECA cluster at JSC~\cite{jureca}, comparing space-parallel, time-serial MLSDC to space-time parallel PFASST runs.
For structured benchmarking, we coupled \pysdc{} to the JUBE Benchmarking Environment~\cite{Lhrs:808798}.
Accuracies, tolerances as well as the code itself are the same for all runs shown, the only difference lies in the distribution of MPI ranks in space and time.
In Figure~\ref{fig:petsc_matrix}, runtimes of different distributions of cores in space and time on a single JURECA compute node are shown.
We can see on the first horizontal line of blocks the purely space-parallel MLSDC runs, ending with $50.6$ seconds for the full simulation.
In contrast, using only $4$ cores in space and $6$ cores in time yields a runtime of $38.6$ seconds, about $1.3$-times faster.
In Figure~\ref{fig:petsc_speedup} we see the time-serial, space-parallel MLSDC run (blue curve, circle markers) up to $24$ cores on one node, which is now extended using time-parallelism (red curve, square markers) to provide speedup on $24$ nodes, i.e.\ on $24\times24$ cores.

All runs have been done using a single \pysdc{} implementation of the heat equation using PETSc for the spatial components. 
The same code runs on all platforms as long as the relevant packages are available. 
We note that porting the code to this HPC cluster was extremely easy: after loading relevant modules on JURECA (in particular the conveniently pre-installed PETSc, petsc4py and mpi4py) we just installed remaining python packages locally via pip and cloned \pysdc{}'s Github repository.
No further installation or adaptation of the code was necessary.
Source files including JUBE templates and job scripts for JURECA can be found in the \texttt{projects/TOMS} folder of \pysdc{}~\cite{pySDC_release}.
\section{Conclusion and outlook}

In this paper we have describe the Python code \pysdc{}, which can be used to prototype spectral deferred corrections (SDC) derived methods like multi-level deferred corrections (MLSDC) and in particular the parallel full approximation scheme in space and time (PFASST).
The main features of \pysdc{} can be summarized as follows:
\begin{itemize}
	\item available implementations of many variants of SDC, MLSDC and PFASST,
    \item many ordinary and partial differential equations already pre-implemented,
    \item seven tutorials to lower the bar for new users and developers,
    \item emulated or MPI-based time-parallelism,
    \item coupling to FEniCS and PETSc, including spatial MPI parallelism for the latter,
    \item automatic testing of new releases, including results of previous publications,
    \item fully compatible with Python 2.7 and 3.6 (at least), runs on desktops and HPC machines.
\end{itemize}

Many of these aspects have been addressed in this paper. 
In particular, we shed light on the availability and structure of the code, including the continuous integration environment used to ensure reliable operability and reproducibility.
Then, two perspectives -- the user's and the developer's perspectives -- have been taken to provide information on the code's functionality.
As a user, the relevant parts of \pysdc{} are the tutorials, which take the user in seven different steps from a simple collocation problem to a full-fledged parallel PFASST run, and the projects, which are more or less complex codes used for publications or for demonstration purposes.
Both parts provide a rather rich, well-documented and tested resource for starting own projects with \pysdc{}.
As a method or algorithm developer, more information on the implementation of \pysdc{} is needed. 
The continuous integration environment automatically generates an API for the core methods, the implementations and a number of smaller helper scripts.
For this paper, we described the way from a front-end script to the actual implementation of the method and the output the controller returns.
Finally, three examples highlighted different aspects of \pysdc{}, in particular the ease of implementation, execution and evaluation of different SDC-related algorithms, the coupling to powerful external libraries like FEniCS or PETSc as well as the ability to even prototype space-time parallel algorithms using MPI.

Not surprisingly, the list of further directions and open tasks is rather long. 
In the following, we highlight parts of this list in four short paragraphs:

\textbf{Cleanup and smoothing} -- Despite a lot of effort, there are of course quite some quirks and rough edges which would need attention at some point. 
One is the somewhat inconsistent handling of classes in the front-end scripts (such as in Listing~\ref{lst:pysdc_main}): for example, historically, it seemed to be a good idea to have collocation classes as part of sweeper parameters, but in hindsight this choice is unexpected for the user and inflexible for the developer. 
Also, problem implementations are crucially tied to data types anyway, so it would make sense to move the specification of data types from the front-end script into the problem classes. 
Another annoyance is the presence of (at least) four different controllers which do more or less the same thing. 
Again, historically this may be understandable, but at the current state of the code it would be more convenient to introduce more abstraction to avoid code replication.
Yet, smoothing out all these (and probably many more) rough edges is desirable, sometimes rather invasive, but always time-consuming.

\textbf{Jupyter notebooks} -- The tutorials and projects shipping with \pysdc{} are currently a collection of Python scripts augmented by a short documentation written in reStructuredText, which also includes plots or other results.
A more user-friendly approach is to join these items into dedicated Jupyter notebooks~\cite{Kluyver:2016aa}, which contain code snippets, documentation, explanation and results and which can be downloaded and executed directly.
In addition to the porting effort, this task involves modification of the current testing environment and deployment mechanism.

\textbf{Parareal} -- With PETSc accessible via petsc4py, an interesting further direction is to provide a Parareal controller within \pysdc{}, using PETSc's ODE integrators as basis.
Most of the structure needed for implementing Parareal is already part of \pysdc{}, e.g.\ the notion of different levels, iterations, advancing in time and different problem classes.
This would be augmented by PETSc's TS class, which contains a multitude of standard ODE solvers.
Ideally, PFASST and Parareal controllers would work with the same problem classes, so that the actual implementation of the ODE or PDE would be exactly the same.
This would allow a direct and fair comparison between both parallel-in-time methods, being based on the same code base, language and implementation.
This also includes a comparison between SDC and standard ODE solvers, which is still rarely seen in the literature.

\textbf{More bindings} -- \pysdc{} already supports data types and solvers from NumPy, FEnICS and PETSc, but there are many more interesting and promising libraries out there, which could help users to implement their applications more easily. 
In particular, we mention here PyClaw~\cite{pyclaw-sisc} and Firedrake~\cite{Rathgeber:2016:FAF:2988516.2998441}, which both provide a lot of domain-specific algorithms and have a strong user community. 
We finally mention that parallelization in space with FEniCS has not been done within \pysdc{} and is left for future work, too.

\bibliographystyle{ACM-Reference-Format}
\bibliography{references}

\end{document}